\documentstyle[11pt,newpasp,twoside,epsf]{article}

\def\edcomment#1{\iffalse\marginpar{\raggedright\sl#1\/}\else\relax\fi}
\reversemarginpar

\def\Pdot{\mathaccent 95 P}
\def\lax{\>\vcenter{\hbox{$<$\hskip-.75em\lower1.0ex\hbox{$\sim$}}}\>}
\def\gax{\>\vcenter{\hbox{$>$\hskip-.75em\lower1.0ex\hbox{$\sim$}}}\>}

\begin{document}

\title{The Spin-Up Rate of the White Dwarf of GK Per}
\author{Christopher W.\ Mauche}
\affil{Lawrence Livermore National Laboratory,\\
L-473, 7000 East Avenue, Livermore, CA 94550}

\begin{abstract}
We use hard X-ray light curves measured by the {\it Chandra\/} HETG and {\it
RXTE\/} PCA during the late rise and plateau phases of the 2002 March--April
outburst of the intermediate polar GK~Per to determine that its X-ray pulse
period $P=351.332\pm 0.002$ s. Combined with previous X-ray and optical
measurements of the spin period of the white dwarf, we find that its spin-up
rate $\Pdot = 0.00027\pm 0.00005~\rm s~yr^{-1}$.
\end{abstract}

\section{Introduction}

GK Per is an extraordinary cataclysmic variable: the first nova of the last
century (Nova Per 1901), it contains a magnetic white dwarf, a large truncated
accretion disk, and an evolved secondary; it is surrounded by a nova shell
visible in the radio, optical, and X-ray wavebands; and it suffers
large-amplitude ($\Delta V\approx 3$), long ($\delta t\approx 60$ days), but
infrequent ($\Delta t\approx 900$ day) dwarf nova outbursts (\v Simon 2002).
On the strength of proposals submitted to the {\it XMM\/} and {\it Chandra\/}
Guest Observer programs, we were granted target-of-opportunity {\it XMM\/} and
{\it Chandra\/} (C.\ Mauche, PI) and {\it RXTE\/} (K.\ Mukai, PI) observations
of GK~Per during its 2002 March--April outburst with the goals of monitoring
the X-ray light curve, pulse profile, and spectrum and obtaining the first
grating spectra of GK Per to determine the nature of its complex X-ray spectrum
(\c Sen \& Osborne 1998; Ezuka \& Ishida 1999). The number and dates of
the observations were constrained by the small and decreasing angle between
GK~Per and the Sun, but we were fortunate to obtain {\it RXTE\/} monitoring
observations every few days throughout the outburst, an {\it XMM\/} observation
on March 9 during the early rise to maximum, and {\it Chandra\/} HETG
observations on March 27 and April 9 during the late rise and peak of the
outburst. The {\it XMM\/} MOS and {\it Chandra\/} HETG confirm that the Fe 6.4
keV fluorescent emission line is the strongest of the three Fe K-shell emission
features, and the {\it XMM\/} RGS and {\it Chandra\/} HETG spectrometers reveal
emission lines of H-like C; H- and He-like N, O, Ne, Mg, Si, and S; and
possibly He-like Al (see Mukai et al.\ 2003 for a plot of the {\it Chandra\/}
HETG spectrum). At least two sources of X-ray emission are implied by (1) the
strong absorption of the continuum (which diminishes to zero for $\lambda\gax
6$~\AA ), (2) the presence of emission lines down to $\lambda \approx 30$~\AA ,
and (3) the fact that the short-wavelength ($\lambda=1$--6~\AA ) flux is
modulated at the spin period of the white dwarf, while the long-wavelength
($\lambda = 6$--26~\AA ) flux is steady.

%Fig1%%%%%%%%%%%%%%%%%%%%%%%%%%%%%%%%%%%%%%%%%%%%%%%%%%%%%%%%%%%%%%%%%%%%%%%
\begin{figure}
\plotone{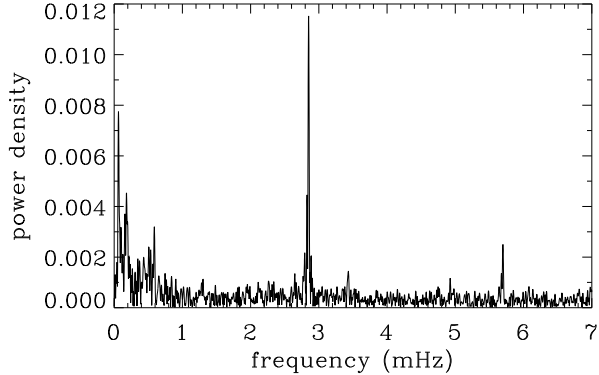} % {1.8in}{0}{100}{100}{-108}{0}
\caption{Power spectrum of the {\it Chandra\/} HETG first-order 1--6~\AA \
light curve of GK~Per. The peaks at $\nu = 2.85$ and 5.68 mHz are identified
with the fundamental and first harmonic of the white dwarf spin 
period $P=351$.}
\end{figure}
%%%%%%%%%%%%%%%%%%%%%%%%%%%%%%%%%%%%%%%%%%%%%%%%%%%%%%%%%%%%%%%%%%%%%%%%%%%%

\section{Spin Period of the White Dwarf}

To determine the spin period of the white dwarf of GK~Per, we used data from
the two 32 ks {\it Chandra\/} observations separated by 13.2 days.
Specifically, we created a count rate light curve of the {\it Chandra\/} HETG
first-order $\lambda=1$--6~\AA \ event data with 351.3/5=70 s bins, subtracted
the mean of the two observations, and padded the interval between the two
observations with zeros. The power spectrum of the resulting light curve is
shown in Figure~1, with peaks at $\nu = 5.68$, 2.85, 0.59, and 0.18 mHz,
corresponding to periods $P = 176$, 351, 1700, and 5520 s; the 351 s period is
the spin period of the white dwarf $P$, the 176 s period is its first harmonic
$P/2$, and the 5520 s period is the quasi-periodic oscillation discussed by
Morales-Rueda, Still, \& Roche (1999). Based solely on the power spectrum, the
spin period of the white dwarf can be constrained only to lie in the range $P=
349$--353 s, but by phasing the two {\it Chandra\/} HETG light curves, the
number of cycles between the two observations is constrained to be 3250 $\pm $
a few, corresponding to periods $P = 351.332$ s $\pm $ multiples of 0.105 s.
This ambiguity is resolved by the light curves of the two intervening {\it
RXTE\/} PCA observations, which phase consistently with the two {\it Chandra\/}
HETG light curves only for $P = 351.332\pm 0.002$ s (see Fig.~2).

%Fig2%%%%%%%%%%%%%%%%%%%%%%%%%%%%%%%%%%%%%%%%%%%%%%%%%%%%%%%%%%%%%%%%%%%%%%%
\begin{figure}
\plotone{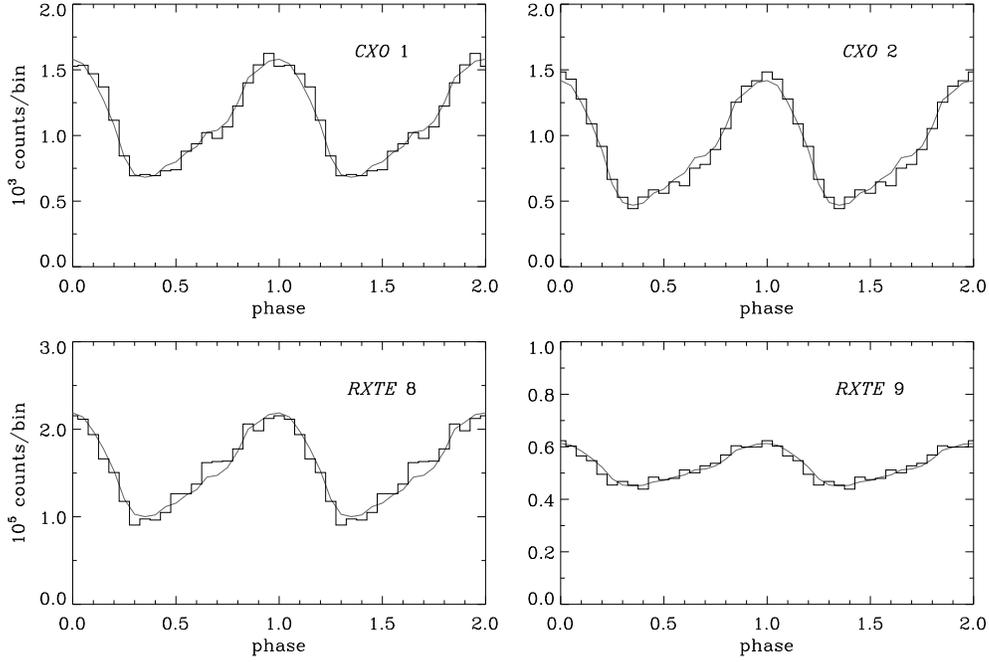} % {2.5in}{0}{100}{100}{-108}{0}
\caption{Phase-folded X-ray light curves of the first and second {\it 
Chandra\/}
HETG and the intervening eighth and ninth {\it RXTE\/} PCA observations of
GK~Per. Data are shown by the histograms and the scaled mean phase-folded light
curve is shown by the gray curves. Phase zero corresponds to HEJD 
2452360.9113.}
\end{figure}
%%%%%%%%%%%%%%%%%%%%%%%%%%%%%%%%%%%%%%%%%%%%%%%%%%%%%%%%%%%%%%%%%%%%%%%%%%%%

%Fig3%%%%%%%%%%%%%%%%%%%%%%%%%%%%%%%%%%%%%%%%%%%%%%%%%%%%%%%%%%%%%%%%%%%%%%%
\begin{figure}
\plotone{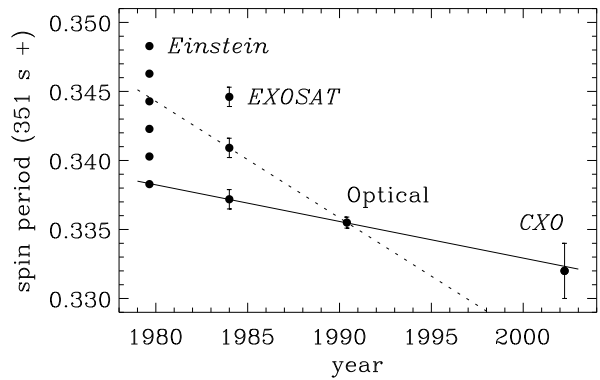} % {1.8in}{0}{100}{100}{-108}{0}
\caption{Spin period history of GK~Per. The solid line is the spin-up rate
$\Pdot= 0.00027~\rm s~yr^{-1}$ consistent with the {\it Einstein\/}, {\it
EXOSAT\/}, optical, and {\it Chandra\/} and {\it RXTE\/} data, and the dotted
line is the value $\Pdot= 0.00084~\rm s~yr^{-1}$ proposed by Patterson (1991).}
\end{figure}
%%%%%%%%%%%%%%%%%%%%%%%%%%%%%%%%%%%%%%%%%%%%%%%%%%%%%%%%%%%%%%%%%%%%%%%%%%%%

\section{Spin-Up Rate of the White Dwarf}

The pulse period of GK~Per has been measured previously in X-rays by {\it
Einstein\/} (Eracleous, Patterson, \& Halpern 1991) and {\it EXOSAT\/} (Watson,
King, \& Osborne 1985; Norton, Watson, \& King 1988) and in the optical by
Patterson (1991), who provides an excellent discussion of the measurements and
uncertainties of the various determinations of this quantity. Patterson finds
that the 1980 {\it Einstein\/} data are consistent with pulse periods $P =
351.3423$ s $\pm $ multiples of 0.002 s, the 1983--85 {\it EXOSAT\/} data are
consistent with pulse periods $P = 351.3372$, 351.3409, or 351.3446 ($\pm
0.0007$) s, and the 1989--91 optical data are consistent with a pulse period
$P=351.3355\pm 0.0004$ s; these data are plotted in Fig.~3. Assuming that the
pulse period at the epoch of the {\it EXOSAT\/} observations was 351.3409 s,
Patterson proposed a spin-up rate $\Pdot = 0.0008~\rm s~yr^{-1}$ (the dotted
line in Fig.~3) for the white dwarf of GK~Per, but cautioned that the result
should be regarded as tentative, as there is an uncertainty in the cycle count.
As shown in Figure~3, our determination of the pulse period of GK~Per in 2002
is not consistent with a linear extrapolation of the spin-up rate proposed by
Patterson, but is consistent with a spin-up rate $\Pdot = 0.00027\pm
0.00005~\rm s~yr^{-1}$ (the solid line in Fig.~3).

\section{Spin Ephemeris of the White Dwarf}

Given the results from the previous sections, the spin ephemeris of the white
dwarf of GK Per is fully determined by specifying the date of some feature
in the light curve. Following Watson, King, \& Osborne (1985), we define
phase zero to occur at pulse maximum, which, guided by the power spectrum, we
determined by fitting a function $a+b \sin [2\pi (\phi -\phi_1)] +c \sin [4\pi
(\phi - \phi_2)]$ to the mean phase-folded light curve. By this method, we
find that X-ray pulse maximum of GK~Per follows the ephemeris $T_{\rm max} =
{\rm HEJD}\, 2452360.9113\, (\pm 0.0001) +0.00406634\, (\pm 0.00000002)\, E -
(3.5\pm 0.6)\times 10^{-14}\, E^2.$ This ephemeris includes only the second
secure determination of the spin period of the white dwarf of GK Per and the
first secure determination of its spin-up rate.

\acknowledgments

This work benefited from discussions with J.\ Patterson, K.\ Mukai, and A.\
Norton.
We thank J.~Swank and H.~Tananbaum for the grants of Director's Discretionary
Time which made the {\it RXTE\/} and {\it Chandra\/} observations possible.
Support for this work was provided in part by NASA through {\it Chandra\/}
Award Number DD2-3014X issued by the {\it Chandra\/} X-Ray Observatory Center,
which is operated by the Smithsonian Astrophysical Observatory for and on
behalf of NASA under contract NAS8-39073. This work was performed under the
auspices of the U.S.~Department of Energy by University of California Lawrence
Livermore National Laboratory under contract No.~W-7405-Eng-48.

% References
%%%%%%%%%%%%%%%%%%%%%%%%%%%%%%%%%%%%%%%%%%%%%%%%%%%%%%%%%%%%%%%%%%%%%%%%%%%%


\begin{references}

\reference Eracleous, M., Patterson, J., \& Halpern, J. 1991, \apj , 370, 330
\reference Ezuka, H., \& Ishida, M. 1999, \apjs , 120, 277
\reference Morales-Rueda, L., Still, M.~D., \& Roche, P. 1999, \mnras , 306,
  753
\reference Mukai, K., et al. 2003, \apj , in press (astro-ph/0301557)
\reference Norton, A.~J., Watson, M.~G., \& King, A.~R. 1988, \mnras , 231, 783
\reference Patterson, J. 1991, \pasp , 103, 1149
\reference \c Sen, G., \& Osborne, J.~P. 1998, in Wild Stars in the Old West,
  ed.\ S.~Howell, E.~Kuulkers, \& C.~Woodward (San Francisco: ASP), 463
\reference \v Simon, V. 2002, \aap , 382, 910
\reference Watson, M.~G., King, A.~R., \& Osborne, J. 1985, \mnras , 212, 917
\end{references}
\end{document}